\documentclass{article}
\usepackage[numbers]{natbib}

\usepackage[final]{neurips_2024}

\usepackage[utf8]{inputenc} % allow utf-8 input
\usepackage[T1]{fontenc}    % use 8-bit T1 fonts
\usepackage{hyperref}       % hyperlinks
\usepackage{url}            % simple URL typesetting
\usepackage{booktabs}       % professional-quality tables
\usepackage{amsfonts}       % blackboard math symbols
\usepackage{nicefrac}       % compact symbols for 1/2, etc.
\usepackage{microtype}      % microtypography
\usepackage{xcolor}    
\usepackage{afterpage}% colors
\usepackage{float}

\title{Composers' Evaluations of an AI Music Tool: Insights for Human-Centred Design}

\author{%
  Eleanor Row \\
  Centre for Digital Music\\
  Queen Mary University of London\\
  London, UK\\
  \texttt{e.r.v.row@qmul.ac.uk} \\
   \And
   Gy{\"o}rgy Fazekas \\ 
   Centre for Digital Music\\
   Queen Mary University of London\\
   London, UK\\
   \texttt{george.fazekas@qmul.ac.uk} \\
}

\begin{document}

\maketitle

% ****ABSTRACT****
\begin{abstract}

We present a study that explores the role of user-centred design in developing Generative AI (GenAI) tools for music composition. Through semi-structured interviews with professional composers, we gathered insights on a novel generative model for creating variations, highlighting concerns around trust, transparency, and ethical design. The findings helped form a feedback loop, guiding improvements to the model that emphasised traceability, transparency and explainability. They also revealed new areas for innovation, including novel features for controllability and research questions on the ethical and practical implementation of GenAI models.
\end{abstract}

% ****PAPER START****
\section{Introduction}
\label{introduction}

Generative AI (GenAI) in creative fields faces challenges due to a divide between human-centric and model-centric research approaches, reflecting a broader disconnect between AI researchers and creative professionals overall, which we highlight in the music domain \cite{newmanHUMANAIMusicCreation2023,vankaAdoptionAITechnology2023, ben-talHowMusicAI2021}. Concerns about trust, ethics, and usability stress a need for interdisciplinary dialogue, which can be used to help guide the development of GenAI tools that meet real-world needs.
To address this in the music domain, we conducted a qualitative study with professional composers, exploring how generative music models can serve as creative aids in music composition. Through semi-structured interviews, composers interacted with a baseline transformer model designed to create musical variations \cite{rowJAZZVARDatasetVariations2023, rowAdvancingAIMusic2024}, providing feedback on outputs, evaluation metrics, transparency, and ethical concerns. This approach established a feedback loop that helps to inform GenAI tool design and foundational research questions, aligning AI development with the practical and ethical needs of end-users. Although centred on music, our findings offer valuable insights for broader applications of GenAI in creative contexts.

\section{Methodology} 
\label{methodology}

Our methodology employed a User-Centred and Participatory Design-based approach \cite{beyerContextualDesignDefining1998, schulerParticipatoryDesignPrinciples1993, hunterUsingParticipatoryDesign2019, fiebrinkMachineLearningAlgorithm2018, mullerParticipatoryDesign1993a}. Four professional composers participated in semi-structured interviews, where they were introduced to a novel music task, Music Overpainting, using a transformer model designed to generate jazz piano variations from basic melodies and chords \cite{rowAdvancingAIMusic2024, rowJAZZVARDatasetVariations2023}. Adapted from the music transformer architecture proposed by Huang et al. \cite{huang2018music}, this model serves as a task-specific baseline for Music Overpainting. Composers could listen to the model's inputs and outputs as much as needed to familiarise themselves with its capabilities. We asked targeted questions for their opinions on how to define a variation, the success of the model, trust and ethical concerns, user integration and controllability aspects as detailed in Appendix \ref{appendix:questions}. These questions helped to address critical aspects of the model's development that were not fully covered by existing literature, aiming to fill knowledge gaps and reduce researcher bias, particularly given the novelty of the generative task. Responses were analysed by one coder, using a reflexive thematic analysis-inspired approach \cite{clarkeSuccessfulQualitativeResearch2013, braunCollectingQualitativeData2017, braunMisConceptualisingThemes2016}. To minimise biases, the coder engaged in regular reflexive practices, including keeping a reflective journal and self-reflecting on biases that they brought in as a researcher. Additionally, we asked the composers to complete a short questionnaire based on \cite{mullensiefenMeasuringFacetsMusicality2014}, collecting demographics and information on their musical education and experience as seen in Appendix \ref{appendix:questionnaire}. While the small sample size and focus on a specific generative task lead to context-specific findings that may not generalise to broader creative communities or demographics, this study serves as an initial exploration that underscores the value of engaging directly with end-users to identify potential sources of researcher bias and enrich GenAI model development, with insights that may otherwise be overlooked. As a preliminary qualitative study, future research could build upon these findings to enhance generalisability by expanding the outputs to additional musical genres and involving a larger and more diverse participant pool.

\section{Findings and Discussion} 
\label{findings}

The key findings from engaging with professional composers, along with the considerations that influenced the model’s development, are summarised in Table \ref{tab:findings}. This table highlights the main themes identified during the study, the specific feedback from composers (see example transcripts in Appendix \ref{appendix:transcripts}), and the resulting design decisions and adjustments made to the model.

\begin{table}[h]
  \caption{Summary of key findings and considerations}
  \label{tab:findings}
  \centering
  \small % Reduce font size
  \begin{tabular}{p{3cm} p{5cm} p{5cm}}
    \toprule
    \textbf{Theme} & \textbf{Findings} & \textbf{Considerations} \\
    \midrule
    Definition of Variation & Composers defined variations as maintaining a recognisable connection to the source, regardless of transformation. & Helped to further define the novel generative task. Reframed evaluation metrics to focus on traceability, emphasising practical relevance over objectivity. \\
    
    Traceability and Transparency & Composers valued transparency in how variations evolve from the source material. & Changes made to the sampling technique, enabling adjustments to the model and user interface (UI) to visually represent transformation steps. \\
    
    Ethical and Trust Concerns & Concerns over usable data i.e. trained with consent and sources were traceable. & Questions to consider on model training data, its source, and how useful it is.  \\
    
    Applicability of Outputs & Composers found traditional subjective evaluations insufficient, preferring assessments of applicability to their workflows. & Reshaped evaluation questions to focus on how outputs fit within the composers' creative processes, shifting from “Is it good?” to “Is it useful?” \\
    
    Practical Model Design & Highlighted practical needs such as controllable, lightweight models that integrate well into existing workflows. & Influenced decisions on model architecture and parameter size, sampling techniques, and UI considerations for an initially simple design that later advances. \\
    \bottomrule
  \end{tabular}
\end{table}

We used composer feedback to improve the same generative variation model introduced in the study. Key concerns around trust, transparency, and ethical design were raised. The findings helped to clarify the generative task which had been defined through researcher bias, and exposed gaps in existing evaluation metrics for new creative tasks, revealing a need for subjective and objective criteria that reflect real-world creative practices. Feedback also emphasised traceability around data collection and model outputs, and explainability as essential for building trust and aligning AI outputs with artistic intentions \cite{bryan-kinnsExploringXAIArts2022, bryan-kinnsUsingIncongruousGenres2024}.

While many of the themes identified in Table \ref{tab:findings} align with domain-general principles from fields like Human-Centred AI (HCAI)— such as interpretability, explainability, and data transparency \cite{weldChallengeCraftingIntelligible2019}  — the direct feedback from composers added a complementary, application-specific perspective, and helped to reveal how the straightforward application of domain-general principles may overlook subtle requirements unique to certain creative tasks. For instance, composers expressed a strong desire for traceability, viewing it not only as a means to track how variations related to, and evolved from the source material but also as a way to support their creative choices. Some suggested that having the ability to visually and interactively follow how the model transforms the output, as a transparent traceable process, could help them better understand and select variations that align with their artistic intentions, providing them with greater control over their creative decisions.
Likewise, the call for controllable and lightweight models reflected not only a general design preference but a specific need for adaptable tools that fit seamlessly into the composers’ iterative workflows. 
Additionally, the study identified new areas to innovate, including the need for novel controllable features. Research questions on how features could be implemented emerged, including approaches to model sampling and tuning. 

Our collaboration helped reduce researcher bias and revealed previously unconsidered user assumptions about creativity and AI’s role as a collaborative partner. It highlighted the need for adaptable, user-responsive AI systems, and encouraged researchers to shift from model-centric to human-centric approaches that prioritise real-world applicability. By incorporating these insights, we can refine model features and interface elements to align AI capabilities with artistic workflows, emphasising a balanced approach where technical functionality is grounded in practical relevance for creative end-users. This in turn fostered interdisciplinary mutual learning by involving composers directly in the development process. Maintaining a continuous feedback loop with creative professionals is crucial, ensuring that AI evolves to meet user needs rather than requiring users to adapt to the AI.

\section{Conclusion and Future Work} 
\label{conclusion}

Looking ahead, extending this approach to other GenAI domains can lay the groundwork for interdisciplinary collaborations that prioritise user-centred AI design. We urge the research community to adopt continuous engagement with end-users, explore innovative features, and focus on designing AI systems that genuinely support and enhance artistic workflows, striking a balance between AI as a creative agent, the needs of the creative community, and the evolving technical landscape of AI development.

\section{Acknowledgements}
The first author is a research student at the UKRI Centre for Doctoral Training in Artificial Intelligence and Music, supported by UK Research and Innovation \textit{[Grant number: EP/S022694/1}].

\bibliographystyle{abbrvnat}

\bibliography{bibliography}

\begin{thebibliography}{18}
\providecommand{\natexlab}[1]{#1}
\providecommand{\url}[1]{\texttt{#1}}
\expandafter\ifx\csname urlstyle\endcsname\relax
  \providecommand{\doi}[1]{doi: #1}\else
  \providecommand{\doi}{doi: \begingroup \urlstyle{rm}\Url}\fi

\bibitem[Ben-Tal et~al.(2021)Ben-Tal, Harris, and Sturm]{ben-talHowMusicAI2021}
O.~Ben-Tal, M.~T. Harris, and B.~L. Sturm.
\newblock How {Music} {AI} {Is} {Useful}: \textit{{Engagements} with {Composers}, {Performers} and {Audiences}}.
\newblock \emph{Leonardo}, 54\penalty0 (5):\penalty0 510--516, Oct. 2021.
\newblock ISSN 0024-094X, 1530-9282.
\newblock \doi{10.1162/leon_a_01959}.
\newblock URL \url{https://tinyurl.com/4bsxa9d4}.

\bibitem[Beyer and Holtzblatt(1998)]{beyerContextualDesignDefining1998}
H.~Beyer and K.~Holtzblatt.
\newblock \emph{Contextual design: defining customer-centered systems}.
\newblock Morgan Kaufmann, San Francisco, Calif, 1998.
\newblock ISBN 978-1-55860-411-7.

\bibitem[Braun and Clarke(2016)]{braunMisConceptualisingThemes2016}
V.~Braun and V.~Clarke.
\newblock ({Mis})conceptualising themes, thematic analysis, and other problems with {Fugard} and {Potts}’ (2015) sample-size tool for thematic analysis.
\newblock \emph{International Journal of Social Research Methodology}, 19\penalty0 (6), June 2016.
\newblock ISSN 1364-5579, 1464-5300.
\newblock \doi{10.1080/13645579.2016.1195588}.
\newblock URL \url{https://tinyurl.com/yc4h6ra5}.
\newblock Publisher: Taylor \& Francis (Routledge).

\bibitem[Braun et~al.(2017)Braun, Clarke, and Gray]{braunCollectingQualitativeData2017}
V.~Braun, V.~Clarke, and D.~Gray, editors.
\newblock \emph{Collecting {Qualitative} {Data}: {A} {Practical} {Guide} to {Textual}, {Media} and {Virtual} {Techniques}}.
\newblock Cambridge University Press, Cambridge, 2017.
\newblock ISBN 978-1-107-05497-4.
\newblock \doi{10.1017/9781107295094}.
\newblock URL \url{https://tinyurl.com/4t4ycd7c}.

\bibitem[Bryan-Kinns et~al.(2022)Bryan-Kinns, Banar, Ford, Reed, Zhang, Colton, Armitage, and Workshop]{bryan-kinnsExploringXAIArts2022}
N.~Bryan-Kinns, B.~Banar, C.~Ford, C.~Reed, Y.~Zhang, S.~Colton, J.~Armitage, and N.~.-e. A. A. f. D. a.~D. Workshop.
\newblock Exploring {XAI} for the {Arts}: {Explaining} {Latent} {Space} in {Generative} {Music}, Dec. 2022.
\newblock URL \url{https://qmro.qmul.ac.uk/xmlui/handle/123456789/77565}.

\bibitem[Bryan-Kinns et~al.(2024)Bryan-Kinns, Noel-Hirst, and Ford]{bryan-kinnsUsingIncongruousGenres2024}
N.~Bryan-Kinns, A.~Noel-Hirst, and C.~Ford.
\newblock Using {Incongruous} {Genres} to {Explore} {Music} {Making} with {AI} {Generated} {Content}.
\newblock In \emph{Creativity and {Cognition}}, pages 229--240, Chicago IL USA, June 2024. ACM.
\newblock ISBN 9798400704857.
\newblock \doi{10.1145/3635636.3656198}.
\newblock URL \url{https://dl.acm.org/doi/10.1145/3635636.3656198}.

\bibitem[Clarke and Braun(2013)]{clarkeSuccessfulQualitativeResearch2013}
V.~Clarke and V.~Braun.
\newblock \emph{Successful {Qualitative} {Research}: {A} {Practical} {Guide} for {Beginners}}.
\newblock Sage, 2013.
\newblock ISBN 978-1-84787-581-5.

\bibitem[Fiebrink and Caramiaux(2018)]{fiebrinkMachineLearningAlgorithm2018}
R.~Fiebrink and B.~Caramiaux.
\newblock The {Machine} {Learning} {Algorithm} as {Creative} {Musical} {Tool}.
\newblock In \emph{The {Oxford} {Handbook} of {Algorithmic} {Music}}. Oxford Handbooks, online edition, Feb. 2018.
\newblock URL \url{https://doi.org/10.1093/oxfordhb/9780190226992.013.23}.

\bibitem[Huang et~al.(2019)Huang, Vaswani, Uszkoreit, Simon, Hawthorne, Shazeer, Dai, Hoffman, Dinculescu, and Eck]{huang2018music}
C.-Z.~A. Huang, A.~Vaswani, J.~Uszkoreit, I.~Simon, C.~Hawthorne, N.~Shazeer, A.~M. Dai, M.~D. Hoffman, M.~Dinculescu, and D.~Eck.
\newblock Music transformer.
\newblock In \emph{International {Conference} on {Learning} {Representations}}, 2019.
\newblock URL \url{https://openreview.net/forum?id=rJe4ShAcF7}.

\bibitem[Hunter et~al.(2019)Hunter, Worthy, Matthews, and Viller]{hunterUsingParticipatoryDesign2019}
T.~Hunter, P.~Worthy, B.~Matthews, and S.~Viller.
\newblock Using {Participatory} {Design} in the {Development} of a {New} {Musical} {Interface}: {Understanding} {Musician}'s {Needs} beyond {Usability}.
\newblock In \emph{Proceedings of the 14th {International} {Audio} {Mostly} {Conference}: {A} {Journey} in {Sound}}, pages 268--271, Nottingham United Kingdom, Sept. 2019. ACM.
\newblock ISBN 978-1-4503-7297-8.
\newblock \doi{10.1145/3356590.3356635}.
\newblock URL \url{https://dl.acm.org/doi/10.1145/3356590.3356635}.

\bibitem[Muller and Kuhn(1993)]{mullerParticipatoryDesign1993a}
M.~J. Muller and S.~Kuhn.
\newblock Participatory {Design}.
\newblock \emph{Communications of the ACM}, 36\penalty0 (6):\penalty0 24--28, June 1993.
\newblock ISSN 0001-0782.
\newblock \doi{10.1145/153571.255960}.
\newblock URL \url{https://doi.org/10.1145/153571.255960}.

\bibitem[Müllensiefen et~al.(2014)Müllensiefen, Gingras, Musil, and Stewart]{mullensiefenMeasuringFacetsMusicality2014}
D.~Müllensiefen, B.~Gingras, J.~Musil, and L.~Stewart.
\newblock Measuring the facets of musicality: {The} {Goldsmiths} {Musical} {Sophistication} {Index} ({Gold}-{MSI}).
\newblock \emph{Personality and Individual Differences}, 60:\penalty0 S35, Apr. 2014.
\newblock ISSN 0191-8869.
\newblock \doi{10.1016/j.paid.2013.07.081}.
\newblock URL \url{https://www.sciencedirect.com/science/article/pii/S019188691300367X}.

\bibitem[Newman et~al.(2023)Newman, Morris, and Lee]{newmanHUMANAIMusicCreation2023}
M.~Newman, L.~Morris, and J.~H. Lee.
\newblock {HUMAN}-{AI} {Music} {Creation}: {Understanding} the perceptions and experiences of music creators for ethical and productive collaboration.
\newblock In \emph{Proceedings of the 24th {International} {Society} for {Music} {Information} {Retrieval} {Conference}}, Nov. 2023.
\newblock URL \url{https://ismir2023program.ismir.net/poster_58.html}.

\bibitem[Row and Fazekas(2024)]{rowAdvancingAIMusic2024}
E.~Row and G.~Fazekas.
\newblock Advancing {AI} in {Music} {Composition}: {Refining} the {Generative} {Music} {Overpainting} {Task}.
\newblock In \emph{{AES} {International} {Symposium} on {AI} and the {Musician}}, Apr. 2024.
\newblock URL \url{https://qmro.qmul.ac.uk/xmlui/handle/123456789/98357}.

\bibitem[Row et~al.(2023)Row, Tang, and Fazekas]{rowJAZZVARDatasetVariations2023}
E.~Row, J.~Tang, and G.~Fazekas.
\newblock {JAZZVAR}: {A} {Dataset} of {Variations} found within {Solo} {Piano} {Performances} of {Jazz} {Standards} for {Music} {Overpainting}.
\newblock In \emph{Proceedings of the 16th {International} {Symposium} on {Computer} {Music} {Multidisciplinary} {Research}}, pages 265--276, Tokyo, Japan, July 2023. Springer.
\newblock \doi{10.5281/zenodo.10113013}.

\bibitem[Schuler and Namioka(1993)]{schulerParticipatoryDesignPrinciples1993}
D.~Schuler and A.~Namioka.
\newblock \emph{Participatory {Design}: {Principles} and {Practices}}.
\newblock CRC Press, Mar. 1993.
\newblock ISBN 978-0-8058-0951-0.

\bibitem[Vanka et~al.(2023)Vanka, Safi, Rolland, and Fazekas]{vankaAdoptionAITechnology2023}
S.~S. Vanka, M.~Safi, J.-B. Rolland, and G.~Fazekas.
\newblock Adoption of {AI} {Technology} in the {Music} {Mixing} {Workflow}: {An} {Investigation}.
\newblock In \emph{{AES} {EUROPE} 2023}, Sept. 2023.

\bibitem[Weld and Bansal(2019)]{weldChallengeCraftingIntelligible2019}
D.~S. Weld and G.~Bansal.
\newblock The challenge of crafting intelligible intelligence.
\newblock \emph{Commun. ACM}, 62\penalty0 (6):\penalty0 70--79, May 2019.
\newblock ISSN 0001-0782.
\newblock \doi{10.1145/3282486}.
\newblock URL \url{https://dl.acm.org/doi/10.1145/3282486}.

\end{thebibliography}

%%%%%%%%%%%%%%%%%%%%%%%%%%%%%%%%%%%%%%%%%%%%%%%%%%%%%%%%%%%%
\newpage
\appendix

\section{Appendix / supplemental material}

\subsection{Questions} \label{appendix:questions}

Participants were asked fifteen semi-structured questions designed to encourage open dialogue and facilitate in-depth discussions. The semi-structured format allowed participants to freely express their thoughts and opinions, fostering a conversational environment. The questions focused on the participants' understanding of variations in music, their perceptions of AI-generated outputs, and potential improvements for AI tools in creative workflows. 

\begin{enumerate}
    \item How would you define a “variation” in the context of music?
    \item On a scale of 1 to 10, how much do you agree with the statement “The outputs generated were successful variations of the original piece?”
    \item Follow-up: Can you explain your rating and what you found successful or unsuccessful about the variations?
    \item Would you consider using a model like this in your music creation process?
    \item Follow-up: If yes, at what stage(s) would you use it - such as improvisation, ideation, or redrafting? If not, why not? 
    \item Do you think using a model like this could enhance or hinder your creative process?
    \item How would you envision this model being integrated into your workflow? 
    \item Follow-up:Would you prefer it as a separate system or integrated into an existing UI? Why?
    \item How would you prefer to interact with this model — through a simple interface with basic controls, or a more advanced system with deep customization options?
    \item Follow-up: What specific features or controls would be most important to you in a user interface for this model?
    \item How much trust would you place in the outputs generated by this model?
    \item Follow-up: What would make you more confident in using AI-generated variations in your work?)
    \item Are there any ethical concerns you have about using AI to generate variations of existing music?
    \item Follow-up: How do you think these concerns could be addressed?
    \item What improvements or changes would you suggest for this model to make it more useful or effective for your needs?

\end{enumerate}

\subsection{Transcripts} \label{appendix:transcripts}

This section provides a selection of transcripts of verbatim responses to questions from interviews conducted with participants. Each participant is labelled anonymously (P01, P02, etc.). The discussions focused on the definition of variations, their impressions of the AI model, and its potential use in music composition.

\begin{list}{}{\setlength{\leftmargin}{1.5cm} \setlength{\labelwidth}{0cm} \setlength{\itemindent}{-1.5cm}}

\item \textbf{Q: How would you define a “variation” in the context of music?}

    \textbf{P01:} I don’t really know. I don’t really know how to describe what a variation of something would be. Like if something is a certain way... I don’t know how you’d... How would you describe like what the variation is? Variation... something that exists and is recognisably the same from the same thing but... or has a recognisable idiom or stylistic similarity to something else. Because the thing is a variation can be anything from one note changing to something almost completely absent from the original idea.

    \textbf{Interviewer:} So there are many different aspects that could be considered varied?

    \textbf{P01:} Yes. So it’s not just the notes not just the harmonies not just the rhythm. Yeah a variation... How close to the original idea it can be... It’s a really tricky definition I think. What is a variation? Because it doesn’t even necessarily have to be recognisably similar to the original. It can be something completely absent. And then you can say "Well it’s a variation of this because there’s one tiny thing about it which has a connection to the original."

    \textbf{Interviewer:} So do you think as long as it has a connection to the original?

    \textbf{P01:} But I mean there’s definitely an argument that even something that doesn’t have any connection to the original... You could say it’s a variation because you can trace back ten different transformations you’ve gone through to get to this point. But you started at an original position but you’ve gone through ten different iterations to get to the variation. But then when you look at the variation it might be completely different to the original.

    \textbf{P02:} I think I would say a variation in the context of music is something that has some sort of core elements of the thing that it is varying but changes it in a way that retains some reference to it but introduces some new ideas. I understand that some people might be like "Oh that’s… it just needs to be a variation and is entirely derived from it." But I think there’s nothing wrong with introducing new ideas in a variation so long as there is a degree of reference to the original one.

    \textbf{P04}: Oh! Well, I mean… it’s just... A repetition. The same thing, but with something that’s different.

    \textbf{P03:} That’s a good question. It’s quite… I feel... like a jazzy technical term I should really know and I don’t. No no just in your own words as a composer. Like adapt some definition for it but I suppose… yeah I guess like an alternative way of playing something or like a different... A reinterpretation or like... Yeah I don’t know. My definitions like... everything that’s springing to mind is about like kind of interpretation or about having some sort of individual like import or transmutation of something. Does that make sense?

\item \textbf{Q: Can you explain what you found successful or unsuccessful about the variations?}

    \textbf{P01:} I mean if we’re using this as a kind of context the point is can we trace what’s actually been done to create these examples? If we can’t then I would say they’re not successful variations of the original idea. If there is a path in which they have been forged... If you know what I mean from the original... then I would say...

    \textbf{P01:} Like I couldn’t necessarily say how the model infers to create those variations. But in terms of if you were to just listen to it... yeah maybe without the traceability aspect... I don’t know. Yeah without the traceability you could sort of infer there’s some kind of similarity. But I think possibly this is an area where... I don’t know whether it’s all AI models but I think there needs to be a kind of rigour put back into analysing where an idea has come from. And I don’t know how possible that is to do with AI models. It comes down to the same thing of... There’s a certain randomness to the output that you don’t necessarily have control over. That worries me slightly because I could not tell you that those are good variations or bad variations because I can’t say whether they are... Other then you know a fleeting resemblance that you might slightly hear. But they’re completely rhythmically harmonically melodically... melodically completely removed from the original ideas as far as I can hear myself.

    \textbf{Interviewer:} So there’s no sort of relation at all?

    \textbf{P01:} Not that I can hear myself but I would be interested if there was some kind of breakdown of how it got to those ideas. But I don’t know how possible that would be to get.

    \textbf{P02:} Even though the first variation was slower I think that harmonically it felt closer to the original. The chords were sort of closer and it had a similar tonality to it. The second one started to introduce a lot of leading notes and chromatic transitions between them which were very much not present in the first one. I can understand why some might say that fits a variation but maybe that was a bit too much of a variation in my mind.

    \textbf{P03:} Like... I suppose... they felt... I suppose first what I found about them is they felt kind of... like a little bit less intentional or like a bit less... But maybe this is my human bias but they felt less intentional like with less sort of direction or something. But then the flip side of that was what I kind of enjoyed about them was that they were... they were kind of rogue. Like I don’t think they would have been... you know I don’t know much like the variations I would have come up with and that is kind of interesting and a good thing you know what I mean?

\item \textbf{Q: Would you consider using a model like this in your music creation process?}

    \textbf{P01:} I think... It would need to be more predictable and more... You know you had the control over it to actually be related to the original idea. So at the very basic level if it was at the same tempo and time signature that would be one step towards getting closer. If it had fragments of the original melody that would be another step to being closer. I can see there’s like a fleeting harmonic resemblance but even the harmony seemed very non-functional in the examples. If it had all those things if it was just honed enough that you could have control over those factors I think it could be potentially useful.

    \textbf{P02:} Yeah definitely. I think that particularly because a lot of what we do is… the kind of model that my music tends to take is very much ‘ABA’. So if I’ve written the ’A’ section and then I’m like "Play me a ’B’ section" or "Help me write a ’B’ section" I think that would be really helpful.

    \textbf{P03:} Yeah I mean I’m not sure what the application would be for me. But I’d be intrigued to try something that was kind of similar and work with it. Yeah because I suppose... I’m like... Again it’s kind of like this wrongness thing or something that always seems to come back a lot in this chat. Like the... I’ve generally leaned towards... Cause this model is dealing with MIDI data rather than audio data. From the examples I’ve heard in terms of applicability to my own work I felt like models that deal with audio I’ve generally found more interesting because they produced weirder and kind of shittier results.

\item \textbf{Q: What specific features or controls would be most important to you in a user interface for this model?}

    \textbf{P01:} Well I don’t really know how it’s functioning. Maybe a slider of how far away from the original idea it gets and then you could vary the metric variability the tempo variability the rhythmic variability... Yeah maybe those are sliders. Like very basic that you could mess around with a little bit and see how far away from it you could get.

    \textbf{P04:} Well hmm. I think my favourite plugins kind of have a multi-tier thing that it’s like... we’re gonna show you just like macros first so you don’t lose your brains and then once you come to grips with what’s happening here then we’ll show you what’s under the hood.

    \textbf{P03:} Yeah I mean I would definitely be into having lots of sliders to slide around and figure out what’s... I like the idea of being able to tweak your weights inside the model and being able to turn... I think being able to mess with the weights and produce... I would really want that to have enough range that you could pull it into the state of wrongness or extremes. I love the idea of being able to pull it really dissonant or really consonant or really far or close from the original. They’re quite nebulous ideas even...

\end{list}

\subsection{Codebook for Thematic Analysis} \label{appendix:codebook}

The following table provides a codebook for the thematic analysis conducted in the study \cite{braunCollectingQualitativeData2017}, which can be used to help reproduce similar research. Each theme represents key insights derived from participant responses, grouped into broader categories that reflect the participants' concerns and observations.

\begin{table}[h]
  \caption{Codebook for Thematic Analysis}
  \label{tab:codebook}
  \centering
  \small % Reduce font size
  \begin{tabular}{p{4cm} p{8cm} p{0.8cm}}
    \toprule
    \textbf{Theme} & \textbf{Description} & \textbf{Codes} \\
    \midrule
    Transparency, Traceability, and Controllability & The need for clear, traceable, and controllable processes in AI-generated variations to enhance trust and usability. & 6 \\
    \midrule
    Intentionality and Interpretation & The importance of intentionality behind variations and how it influences the perceived quality and relevance of AI outputs. & 4 \\
    \midrule
    Creativity as a Collaborative Process & Framing the model as a creative partner, inspiring new ideas rather than replacing human creativity. & 5 \\
    \midrule
    Customization and Adaptability & A balance between simplicity and depth, allowing users to customize the AI's outputs to better fit their creative needs. & 3 \\
    \midrule
    Predictability and Familiarity & Maintaining a degree of predictability in outputs to ensure they align with the user's expectations and creative intent. & 4 \\
    \bottomrule
  \end{tabular}
\end{table}
\begin{minipage}{\textwidth}
\subsection{Consent and Risk Forms} \label{appendix:consent}
Participants were informed about the study, its purpose, and their rights, including voluntary participation, the ability to withdraw at any time, and the secure, anonymous handling of their data. The consent form outlined the scope of data collection, including screen, audio, and video recordings, and highlighted the data protection measures in place. Participants were required to acknowledge each statement before agreeing to take part in the study. A section from the participant information form on risks is also included below.

\vspace{1em}

\textbf{Title of Research Study:} Composing Culture: A Brief Ethnography of Composers and their Creative Process

\vspace{1em}

\textbf{Principal Investigator:} Eleanor Row

\vspace{1em}

Thank you for your interest in this research. Should you wish to participate in the study, please consider the following statements. Before signing the consent form, you should initial all or any of the statements that you agree with. Your signature confirms that you are willing to participate in this research, however, you are reminded that you are free to withdraw your participation at any time.

\vspace{1em}

\textbf{Ethics of Research Committee Ref:} QMERC20.565.DSEECS23.058 

\vspace{1em}

\textbf{Risk assessment:}
Both the observation and interview sessions will require your dedicated time. While we prioritise your privacy and anonymity, sharing your screen and photographs of workspace involves some level of exposure. Precautions will be taken to ensure
your anonymity. Remember, you always have the choice to opt-out or restrict any part of the process if it makes you feel uneasy. Possible technical issues with Zoom or other tools used during the study might affect the study experience.
\vspace{1em}
\noindent

\textbf{Statement of Consent}

\begin{table}[H]
  \centering
  \small
  \begin{tabular}{p{12cm} p{0.8cm}}
    \toprule
    \textbf{Statement} & \textbf{Initials} \\
    \midrule
    1. I confirm that I have read the Participant Information Sheet dated (retracted) version 1.01 for the above study; or it has been read to me. I have had the opportunity to consider the information, ask questions and have had these answered satisfactorily. & \\
    
    2. I understand that my participation is voluntary and that I am free to stop taking part in the study at any time without giving any reason and without my rights being affected. & \\
    
    3. I understand that my data will be accessed by the principal investigator Eleanor Row. & \\
    
    4. I understand that collected data is completely anonymous and information that I have provided cannot be withdrawn after submission. & \\
    
    5. I understand that my data will be securely stored in the United Kingdom, and in accordance with the data protection guidelines of Queen Mary University of London for 5 years in fully anonymous form. & \\
    
    6. I understand that the information collected about me will be used to support other research in the future, and it may be shared in anonymous form with other researchers. & \\
    
    7. I understand that my shared screen will be recorded and that my audio and video will also be recorded via Zoom. & \\
    
    8. I agree to take part in the above study. & \\
    \bottomrule
  \end{tabular}
\end{table}

\end{minipage}

\textit{Participants should read the QMUL privacy notice for research participants, which contains important information about your personal data and your rights in this respect. If you have any questions relating to data protection, please contact the Data Protection Officer.}

\begin{minipage}{\textwidth}
\vspace{2em}
\subsection{Questionnaire Responses} \label{appendix:questionnaire}

This appendix provides a detailed summary of the responses collected from the participant questionnaire. The questionnaire was designed to gather demographic information, and information on composers' musical backgrounds and experiences with questions stemming from Goldsmith's Musical Sophistication Index (Gold MSI) \cite{mullensiefenMeasuringFacetsMusicality2014}. 

\end{minipage}
\begin{table}[H]
  \caption{Demographic Information and Musical Education of Participants}
  \label{tab:demographics_education}
  \centering
  \small
  \begin{tabular}{p{3cm} p{5cm} p{4.5cm}}
    \toprule
    \textbf{Question} & \textbf{Responses} & \textbf{Details} \\
    \midrule
    Age & 25-34 & All participants were in the 25-34 age range. \\
    
    Ethnicity & White, Mixed/Multiple Ethnic Groups & Majority identified as White, with one identifying as Mixed/Multiple Ethnic Groups. \\
    
    Sexual Orientation & Heterosexual (straight), Prefer not to answer, Queer, Homosexual (lesbian or gay) & Participants expressed varied sexual orientations, including Queer and Heterosexual. \\
    
    Gender Identity & Man, Genderqueer/gender fluid, Woman & Participants identified primarily as men, with one identifying as a woman and another as genderqueer/gender fluid. \\
    
    Disability Status & No, Yes & One participant reported having a disability. \\
    
    Occupation & Composer, AV Technician, Tutor, PhD Student & Majority were composers, with other roles including AV Technician and Tutor. \\
    
    Highest Educational Qualification & Postgraduate degree, Undergraduate degree & Most participants had postgraduate qualifications, with one still pursuing education. \\
    
    Highest Qualification Expected (if still in education) & Not applicable, Postgraduate degree, PhD & Relevant to those still studying; one pursuing a PhD. \\
    
    Music-Related Qualifications & ABRSM/Trinity Grade 5+, Postgraduate (Masters), Undergraduate, GCSE & Participants had various music-related qualifications, from ABRSM/Trinity to postgraduate degrees. \\
    \bottomrule
  \end{tabular}
% \end{table}
  \vspace{1em} % Add spacing between tables
  \caption{Summary of Participant Questionnaire Responses based on Gold-MSI \cite{mullensiefenMeasuringFacetsMusicality2014}}
  \label{tab:questionnaire_responses}
  \centering
  \small

  % % \afterpage{
  % \begin{table}
  % \caption{Table 4: Summary of Participant Questionnaire Responses based on Gold MSI \cite{mullensiefenMeasuringFacetsMusicality2014}}
  % \label{tab:questionnaire_responses}
  % \centering
  % \small
  \begin{tabular}{p{3cm} p{5cm} p{4.5cm}}
    \toprule
    \textbf{Question} & \textbf{Response} & \textbf{Details} \\
    \midrule
    Complimented for musical talents & Neither Agree nor Disagree, Disagree, Completely Disagree & Participants mostly did not feel recognised for their musical talents. \\
    
    Consider as a musician & Strongly Disagree, Neither Agree nor Disagree, Disagree, Completely Disagree & Most participants did not consider themselves musicians, with varying degrees of disagreement. \\
    
    Years of regular practice & 4-5, 2, 0 & Ranged from no regular practice to up to 5 years of daily practice. \\
    
    Hours practised daily & 5 or more, 1, 3-4, 1 & Practice ranged from an hour a day to over 5 hours at peak interest. \\
    
    Instruments played & 3, 2, 3, 4 & Participants played multiple instruments, reflecting diverse musical backgrounds. \\
    
    Listening time & 1-2 hrs, 15-30 min, 4 hrs or more, 15-30 min & Varied listening habits from minimal to extensive daily music listening. \\
    
    Best instrument & Piano, Voice, Violin, Piano/Keys & Main instruments varied widely among participants, showing different primary musical skills. \\
    
    Years of formal training (Music Theory) & 0, 4-6, 2, 0 & Experience in music theory ranged from none to several years. \\
    
    Years of formal training (Instrument) & 7 or more, 4-6, 1, 4-6 & Participants had varying lengths of formal instrument training, some with extensive experience. \\
    
    Years of formal training (Composition) & 4-6, 4-6, 4-6, 0 & Most had some composition training, with one participant having none. \\
    
    Engagement frequency & Everyday & All participants reported daily engagement in musical activities. \\
    
    Main musical genre & Classical, Other, Rock/Pop, Rock/Pop & Participants' musical preferences spanned all the genres listed. \\
%     \bottomrule
%   \end{tabular}
% \end{table}
% }
    \bottomrule
  \end{tabular}
\end{table}

 % \afterpage{
%%%%%%%%%%%%%%%%%%%%%%%%%%%%%%%%%%%%%%%%%%%%%%%%%%%%%%%%%%%%
% \FloatBarrier
\newpage
\section*{NeurIPS Paper Checklist}

\begin{enumerate}

\item {\bf Claims}
    \item[] Question: Do the main claims made in the abstract and introduction accurately reflect the paper's contributions and scope?
    \item[] Answer: \answerYes{} % Replace by \answerYes{}, \answerNo{}, or \answerNA{}.
    \item[] Justification: The abstract and introduction reflect the main contributions of the paper - a study on user-centred design that emphasises the importance of composer feedback and the broader implications for AI in the creative fields.
    \item[] Guidelines:
    \begin{itemize}
        \item The answer NA means that the abstract and introduction do not include the claims made in the paper.
        \item The abstract and/or introduction should clearly state the claims made, including the contributions made in the paper and important assumptions and limitations. A No or NA answer to this question will not be perceived well by the reviewers. 
        \item The claims made should match theoretical and experimental results, and reflect how much the results can be expected to generalise to other settings. 
        \item It is fine to include aspirational goals as motivation as long as it is clear that these goals are not attained by the paper. 
    \end{itemize}

\item {\bf Limitations}
    \item[] Question: Does the paper discuss the limitations of the work performed by the authors?
    \item[] Answer: \answerNo{} % Replace by \answerYes{}, \answerNo{}, or \answerNA{}.
    \item[] Justification: Yes and No. The paper tries to address some of the limitations of the work within the methodology section (including the sample size and context-specific findings), but due to restrictions of the page limit, it is not able to include all of the limitations of the paper. 

\item {\bf Theory Assumptions and Proofs}
    \item[] Question: For each theoretical result, does the paper provide the full set of assumptions and a complete (and correct) proof?
    \item[] Answer: \answerNA{} % Replace by \answerYes{}, \answerNo{}, or \answerNA{}.
    \item[] Justification: The paper does not include theoretical results.

    \item {\bf Experimental Result Reproducibility}
    \item[] Question: Does the paper fully disclose all the information needed to reproduce the main experimental results of the paper to the extent that it affects the main claims and/or conclusions of the paper (regardless of whether the code and data are provided or not)?
    \item[] Answer: \answerNA{} % Replace by \answerYes{}, \answerNo{}, or \answerNA{}.
    \item[] Justification: The paper does not include experiments.

\item {\bf Open access to data and code}
    \item[] Question: Does the paper provide open access to the data and code, with sufficient instructions to faithfully reproduce the main experimental results, as described in supplemental material?
    \item[] Answer: \answerYes{} % Replace by \answerYes{}, \answerNo{}, or \answerNA{}.
    \item[] Justification: The paper provides open access to the questions, transcripts and also code book for Thematic Analysis employed within the study.

\item {\bf Experimental Setting/Details}
    \item[] Question: Does the paper specify all the training and test details (e.g., data splits, hyperparameters, how they were chosen, type of optimizer, etc.) necessary to understand the results?
    \item[] Answer: \answerNA{} % Replace by \answerYes{}, \answerNo{}, or \answerNA{}.
    \item[] Justification: The paper does not include experiments.

\item {\bf Experiment Statistical Significance}
    \item[] Question: Does the paper report error bars suitably and correctly defined or other appropriate information about the statistical significance of the experiments?
    \item[] Answer: \answerNA{} % Replace by \answerYes{}, \answerNo{}, or \answerNA{}.
    \item[] Justification: The paper does not include experiments.

\item {\bf Experiments Compute Resources}
    \item[] Question: For each experiment, does the paper provide sufficient information on the computer resources (type of compute workers, memory, time of execution) needed to reproduce the experiments?
    \item[] Answer: \answerNA{} % Replace by \answerYes{}, \answerNo{}, or \answerNA{}.
    \item[] Justification: The paper does not include experiments.
    
\item {\bf Code Of Ethics}
    \item[] Question: Does the research conducted in the paper conform, in every respect, with the NeurIPS Code of Ethics \url{https://neurips.cc/public/EthicsGuidelines}?
    \item[] Answer: \answerYes{} % Replace by \answerYes{}, \answerNo{}, or \answerNA{}.
    \item[] Justification: This research conforms to the NeurIPS Code of Ethics. The study underwent an institutional ethics review process, ensuring adherence to ethical standards for research involving human participants. All data collected was anonymised, securely stored on an encrypted disk, and will be deleted after five years. Informed consent was obtained from all participants, with clear communication regarding data usage and privacy protections. Further information on the ethics review process can be found in the Appendix section. 

\item {\bf Broader Impacts}
    \item[] Question: Does the paper discuss both potential positive societal impacts and negative societal impacts of the work performed?
    \item[] Answer: \answerNo{} % Replace by \answerYes{}, \answerNo{}, or \answerNA{}.
    \item[] Justification: Due to the page limit, there was not enough space to discuss the negative societal impacts of the work. Only the positive societal impacts were discussed. 
    
\item {\bf Safeguards}
    \item[] Question: Does the paper describe safeguards that have been put in place for responsible release of data or models that have a high risk for misuse (e.g., pretrained language models, image generators, or scraped datasets)?
    \item[] Answer: \answerNA{} % Replace by \answerYes{}, \answerNo{}, or \answerNA{}.
    \item[] Justification: The paper poses no such risks.

\item {\bf Licenses for existing assets}
    \item[] Question: Are the creators or original owners of assets (e.g., code, data, models), used in the paper, properly credited and are the license and terms of use explicitly mentioned and properly respected?
    \item[] Answer: \answerYes{} % Replace by \answerYes{}, \answerNo{}, or \answerNA{}.
    \item[] Justification: The baseline model that was used in the study was appropriately cited and credited. 

\item {\bf New Assets}
    \item[] Question: Are new assets introduced in the paper well documented and is the documentation provided alongside the assets?
    \item[] Answer: \answerNA{} % Replace by \answerYes{}, \answerNo{}, or \answerNA{}.
    \item[] Justification: The paper does not release new assets.

\item {\bf Crowdsourcing and Research with Human Subjects}
    \item[] Question: For crowdsourcing experiments and research with human subjects, does the paper include the full text of instructions given to participants and screenshots, if applicable, as well as details about compensation (if any)? 
    \item[] Answer: \answerYes{} % Replace by \answerYes{}, \answerNo{}, or \answerNA{}.
    \item[] Justification: Participant instructions are included in the supplemental material.

\item {\bf Institutional Review Board (IRB) Approvals or Equivalent for Research with Human Subjects}
    \item[] Question: Does the paper describe potential risks incurred by study participants, whether such risks were disclosed to the subjects, and whether Institutional Review Board (IRB) approvals (or an equivalent approval/review based on the requirements of your country or institution) were obtained?
    \item[] Answer: \answerYes{} % Replace by \answerYes{}, \answerNo{}, or \answerNA{}.
    \item[] Justification: Ethics approval was received for the study, and potential risks were laid out to applicants in the call for participants (provided in the supplemental material). However, due to space constraints, this was not explicitly mentioned in the paper but in the appendix section. 

\end{enumerate}

\end{document}